\documentclass[conference]{IEEEtran}
\usepackage{float}
\IEEEoverridecommandlockouts

\usepackage{tikz}
\newcommand*\circled[1]{\tikz[baseline=(char.base)]{
            \node[shape=circle,draw,inner sep=0pt] (char) {#1};}}

\IEEEaftertitletext{\vspace{-1.5\baselineskip}}
\usepackage{amsthm}
\usepackage{amsmath,amssymb,amsfonts}
\usepackage{algorithmicx}
\usepackage{algcompatible}
\usepackage[ruled,vlined,linesnumbered]{algorithm2e}
\usepackage{graphicx}
\usepackage{textcomp}
\usepackage{xcolor}
\usepackage{float}
\usepackage{graphicx}
\usepackage{subcaption} 

\usepackage{cite}
\usepackage{url}

\newtheorem{remark}{Remark}

\theoremstyle{definition} 
\newtheorem{definition}{Definition}

\usepackage{etoolbox}
\makeatletter
\patchcmd{\thebibliography}{\section*{\refname}}{%
  \section*{\refname}\vspace{-0.8em}
}{}{}
\makeatother
\begin{document}

\title{
Policy-Guided MCTS for near Maximum-Likelihood Decoding of Short Codes\\
}


\author{
\IEEEauthorblockN{
Yuan Tian$^{1}$, Chentao Yue$^{1}$, Peng Cheng$^{2}$, Gaoyang Pang$^{1}$, Branka Vucetic$^{1}$, and Yonghui Li$^{1}$
}
\IEEEauthorblockA{
$^{1}$School of Electrical and Computer Engineering, The University of Sydney, Australia\\
$^{2}$Department of Computer Science and Information Technology, La Trobe University, Australia\\
Email: \{yuan.tian1, chentao.yue, gaoyang.pang, branka.vucetic, yonghui.li\}@sydney.edu.au, p.cheng@latrobe.edu.au
}
\thanks{The work of Chentao Yue was supported by ARC DECRA under Grant DE250101332. Code available: https://github.com/1014Tea/MCTS-Decoding.}
}

\maketitle

\begin{abstract}
In this paper, we propose a policy-guided Monte Carlo Tree Search (MCTS) decoder that achieves near maximum-likelihood decoding (MLD) performance for short block codes. The MCTS decoder searches for test error patterns (TEPs) in the received information bits and obtains codeword candidates through re-encoding. The TEP search is executed on a tree structure, guided by a neural network policy trained via MCTS-based learning. The trained policy guides the decoder to find the correct TEPs with minimal steps from the root node (all-zero TEP). The decoder outputs the codeword with maximum likelihood when the early stopping criterion is satisfied. The proposed method requires no Gaussian elimination (GE) compared to ordered statistics decoding (OSD) and can reduce search complexity by 95\% compared to non-GE OSD. It achieves lower decoding latency than both OSD and non-GE OSD at high SNRs.

\end{abstract}


\vspace{-0.5em}
\section{Introduction}
\vspace{-0.3em}
Ultra-reliable low-latency communications (URLLC) has been defined by the third generation partnership project (3GPP) as a key 
service categories in beyond-5G and 6G networks \cite{chen20235g}. 
URLLC requires end-to-end latency below 1 ms and block error rate (BLER) of $10^{-5}$–$10^{-7}$~\cite{shirvanimoghaddam2018short}. 
To satisfy these requirements, URLLC will demand short channel codes with low decoding complexity and superior BLER performance \cite{shirvanimoghaddam2018short,yue2023efficient}. Among existing solutions, low-density parity-check (LDPC) codes with belief propagation (BP) decoding offer low complexity, but suffer significant BLER performance degradation at short block length \cite{shirvanimoghaddam2018short}. High-density codes such as Bose-Chaudhuri-Hocquenghem (BCH) codes can approach the finite-length bound under near maximum likelihood decoding (MLD)  \cite{yue2023efficient}; however, the exponential decoding complexity hinders their application. Recently, deep learning (DL) has been applied to channel coding and decoding, with extensive work on enhancing BP decoding via deep neural networks (DNNs), including neural belief propagation (NBP)~\cite{nachmani2019hyper,nachmani2022neural}, as well as joint optimization of codes and BP decoders~\cite{larue2022neural,dufrene2025learning}. These methods show the potential of DL to enhance existing decoders. Nevertheless, BP and its DL-based variants rely on the Tanner graph of the code.  They remain fundamentally limited and exhibit a persistent performance gap from MLD at short block lengths. Short codes typically have dense parity-check matrices that induce short cycles in code Tanner graphs, causing correlation among messages and degraded BP decoding performance.

List-based decoding offers an alternative to achieve near-MLD. These decoders generate a candidate list of codewords and identify the one with the minimum Euclidean distance to the received signal. 
Representatives include Chase decoding \cite{chase2003class} and ordered statistics decoding (OSD)~\cite{Fossorier1995OSD}. OSD re-orders the columns of the code generator matrix based on the reliability of the received bits, then transforms it into systematic form via Gaussian elimination (GE), which places the most reliable bits in the information set. Then it applies test error patterns (TEPs) to the information set 
to obtain codeword candidates. 
An order-$m$ OSD searches $O(k^{m})$ TEPs 
with message length $k$ \cite{yue2025guesswork}. 
Recent work has also proposed non-GE OSD variants\cite{yue2022ordered}, which omit the GE overhead but require examining more TEPs, particularly at low SNRs.

Existing efforts to reduce OSD complexity mainly focus on early stopping criterion \cite{yue2022ordered,wu2006soft,jin2006probabilistic} and discarding unnecessary TEPs \cite{yue2022ordered,liang2022low,li2024order}. 
Nevertheless, decoding efficiency also heavily depends on the TEP search strategy\cite{yue2019segmentation}. Existing decoders typically search TEPs in ascending order of Hamming weight, as fewer errors are more probable. However, this strategy often requires examining many TEPs before finding the correct TEP with large Hamming weight. Monte Carlo Tree Search (MCTS) offers a promising approach to address this TEP search problem. MCTS has achieved remarkable success in complex domains such as Go, where  AlphaGo{\cite{silver2016mastering}} and AlphaZero~\cite{silver2017mastering} reached superhuman performance by combining MCTS with DNNs. In these systems, a learned policy guides the search toward promising regions, while MCTS balances exploration and exploitation through statistical simulations during training and inference.  

In this paper, we develop a policy-guided MCTS decoder that achieves near-MLD performance for short codes. 
Instead of enumerating TEPs by ascending Hamming weight, our proposed method organizes the TEP space as a deterministic binary tree and traverse it using depth-first search. 
A neural network policy, trained via MCTS, guides branch selection during the decoding.
For training, MCTS explores the tree guided by the current policy network, and the statistics from successful trajectories are used 
to refine the policy network iteratively. 
Once trained, the policy directs the search toward the correct TEP along the shortest path.
Experiments show that our approach 
requires significantly fewer TEP searches compared to non-GE OSD. 
For the $(48,24)$ code at 1 dB, our decoder needs fewer than 200 TEPs versus over 5000 for non-GE OSD to find MLD results. It also achieves lower decoding time than OSD at high SNRs, as it avoids GE.
The proposed search framework can be generalized to any OSD variants and work with any existing early stopping or TEP pruning method.

The rest of this paper is organized as follows. Section~\ref{Pre} introduces the preliminaries. Section~\ref{sec::MCTS::decoding} presents the proposed MCTS-guided decoding algorithm. The simulation results are discussed in Section~\ref{sec:results}, and Section~\ref{sec:: conclusion} concludes the paper.

\vspace{-0.5em}
\section{Preliminaries}\label{Pre}
\vspace{-0.3em}
\subsection{Linear Block Codes and Channel Model} \label{sec::model}
\vspace{-0.3em}
A binary linear block code $\mathcal{C}(n,k)$ {is} defined by a generator matrix $\mathbf{G}\in\{0,1\}^{k\times n}$, where $n$ and $k$ denote the length of the codeword and the message, respectively. The codeword $\mathbf{c}$ is obtained as $\mathbf{c}=\mathbf{bG}$, where $\mathbf{b}\in\{0,1\}^k$ is the  binary message. The codeword $\mathbf{c} = [c_1,\ldots,c_n]$ is transmitted over the additive white Gaussian noise (AWGN) channel with binary phase shift keying (BPSK) modulation. Let {$\mathbf{x}=1-2\mathbf{c} \in\{-1,1\}^n$} be the modulated symbol of $\mathbf{c}$ and $\mathbf{z}$ be the AWGN vector with noise power $\sigma^2$, i.e., $z_i \sim \mathcal{N}(0,\sigma^2)$. The received signal $\mathbf{r} = [r_1,\ldots,r_n]$ is obtained as $\mathbf{r=x+z}$. We define the SNR in decibels as
$\gamma = 10\log_{10}\!\left(\frac{1}{\sigma^2}\right)$.


\vspace{-0.3em}
\subsection{Ordered Statistics Decoding {and its non-GE variant}}
\vspace{-0.3em}
OSD provides a practical approach to near-MLD performance. Let log-likelihood ratios (LLRs) for each received bit be
$
    \mathbf{\ell}_i  = \ln\frac{\mathrm{Pr}(c_i=0|r_i)}{\mathrm{Pr}(c_i=1|r_i)}=\frac{2r_i}{\sigma^2},
$
for $i = \{1,\dots,n\}$. OSD performs a column permutation $\pi_{\mathrm{OSD}}$ on the generator matrix $\mathbf{G}$ according to the descending absolute LLR values $|\ell_i|$, such that columns corresponding to more reliable bits are placed in the first $k$ positions. GE is then applied to the permuted generator matrix to obtain a systematic form $\mathbf{G}' = [\mathbf{I}_k \mid \mathbf{P}]$, where $\mathbf{I}_k$ is a $k\times k$ identity matrix that corresponds to the $k$ most reliable bits.
Let ${\mathbf{b_0'}} \in \{0,1\}^k$ denote the hard-decision vector of the $k$ most reliable bits, referred to as the most reliable basis (MRB). A TEP $\mathbf{e}\in\{0,1\}^k$ with Hamming weight $w(\mathbf{e})$ modifies the MRB via $
\mathbf{b}_\mathbf{e}'={\mathbf{b}}_0'\oplus \mathbf{e}$. 
Then, a candidate codeword ${\mathbf c_{\mathbf{e}}}$ is generated by re-encoding $\mathbf{b}_{\mathbf{e}}'$ according to $\mathbf c_{\mathbf{e}}'=\mathbf b_\mathbf{e}'\,\mathbf G{'}${, and applying the inverse permutation $\mathbf c_{\mathbf{e}} = \pi_{\mathrm{OSD}^{-1}}(\mathbf c_{\mathbf{e}}')$}. An order-$m$ OSD searches only TEPs with Hamming weight $w(\mathbf{e}) \leq m$, which are typically enumerated in ascending weight order, as fewer bit errors are more likely \cite{Fossorier1995OSD}. Therefore, the total number of TEPs is $\sum_{i=0}^m\binom{k}{i} = O(k^m)$.
Finally, the decoder outputs the candidate codeword 
with the minimum Euclidean distance $d(\mathbf{c}_\mathbf{e},\mathbf{r})$ to the received signal, where $d(\mathbf{c}_\mathbf{e},\mathbf{r}) = \|\mathbf r-\mathbf (1-2\mathbf{c}_\mathbf{e})\|_2$.

The non-GE OSD variant \cite{yue2022ordered} avoids both the column permutation and Gaussian elimination steps.  Instead, it applies TEPs directly to the hard decisions of the first $k$ received bits and encodes all candidates via the original generator 
$\mathbf{G}$. Specifically, let $\mathbf{b}_{0}$ denote the hard-decision vector of the first $k$ position of $\mathbf{r}$. Non-GE OSD directly obtain a codeword candidates by $\mathbf{c_e} = (\mathbf{b_0}\oplus \mathbf{e)} \mathbf{G}.$ At high SNRs, OSD often finds the correct TEP within a few search steps and can terminate early using stopping rules \cite{yue2022ordered} or cyclic redundancy check (CRC). In this regime, the GE overhead dominates the total cost, making non-GE OSD more efficient \cite{yue2022ordered}. At low SNRs, however, non-GE OSD requires a much higher decoding order $m$ 
to achieve the same BLER as the original OSD, since the fixed information set is less reliable than MRB. Prior work has addressed the low-SNR inefficiency of non-GE OSD through adaptive GE reduction \cite{yue2022ordered}, affine-invariant permutations \cite{yu2025ordered}, and staircase generator matrices \cite{fossorier2024modified}.

In this paper, we employ the policy-guided MCTS to reduce the TEP search complexity in non-GE OSD. Notably, the proposed framework can be readily extended to GE-based OSD and other decoder variants involving TEP search.

\vspace{-0.3em}
\subsection{Monte Carlo Tree Search} \label{sec::MCTS::standard}
\vspace{-0.3em}
Following AlphaZero, MCTS can be viewed as a self-play reinforcement learning procedure \cite{silver2017mastering}.   MCTS iteratively balances exploration and exploitation {during tree search} through four steps \cite{silver2016mastering} (See Fig. \ref{fig:placeholder3}).

\subsubsection{Selection}
Starting from the root, repeatedly choose a child according to a selection policy (e.g., PUCT with DNN \cite{silver2017mastering}{, see Section \ref{sec::MCTS}}), descending the current tree until reaching a node that can be further expanded to children. 

\subsubsection{Expansion}
Upon reaching a node that can be further expanded, add one or a set of new child nodes to the tree, thereby growing the search frontier.

\subsubsection{Simulation (Evaluation)}
From the newly added node, perform a simulation to evaluate the \textit{node value}, by using a neural network or random selection to quickly expand the tree into deeper levels and estimate future potential outcomes.

\subsubsection{Backpropagation}
Propagate the estimated value back along the path to the root, updating the statistics (visit counts and accumulated values) at each visited node. These updated statistics guide future selection steps and policy training.

These four steps are executed iteratively for multiple iterations. 
Unlike AlphaZero, in this paper, we can leverage the ground truth (e.g., MLD codeword) for training. MCTS explores the TEP tree to identify successful search paths, generating high-quality supervisory trajectories to train the policy network. The improved policy subsequently biases MCTS toward correct decoding paths with higher probability in the subsequent training iterations. This iterative process, where the algorithm self-generates training supervision, represents a form of self-supervised learning \cite{8831394}. Unlike conventional supervised learning, MCTS training requires no optimal-path labels, yielding stronger generalization across different received noisy signals (see Section \ref{sec::MCTS::decoding}). 


\vspace{-0.5em}
\section{Policy-Guided MCTS Decoding} \label{sec::MCTS::decoding}

\vspace{-0.3em}
\subsection{Tree Structure of TEPs and MCTS Formulation} \label{sec::tree}
\vspace{-0.3em}
{Inspired by \cite{yue2021probability}}, we build a binary tree for TEPs as follows.

\begin{definition}[TEP Tree] \label{def::TEP}
A TEP tree is a deterministic binary tree that enumerates TEPs $\mathbf{e}\in\{0,1\}^k$ with Hamming weight $w(\mathbf{e})\leq m$, where $1\leq m\leq k$.  Each node corresponds to a TEP $\mathbf{e}$, represented by the strictly increasing index set of its non-zero positions
$$\mathcal{Z}_{\mathbf{e}}=\{z_1,z_2,\ldots,z_\ell\},$$
where $z_1 < z_2 < \cdots < z_\ell$ and $\ell=w(\mathbf{e})$. The root of the tree is defined as the all-zero TEP with $\mathcal{Z}_{\mathbf{0}}=\varnothing$. 
From any node $\mathcal{Z}_{\mathbf{e}}$, at most two children are generated:
\begin{itemize}
    \item Extended child 
$
\mathcal{Z}_{\mathbf{e}}^{\mathrm{ext}}=\{z_1,\ldots,z_\ell,\,k\}.
$

\item Adjacent child $\mathcal{Z}_{\mathbf{e}}^{\mathrm{adj}}=\{z_1,\ldots,z_{\ell-1},\,z_\ell-1\}.$
\end{itemize}

Children are created only when feasible. The extended child requires $z_\ell \neq k$, and the adjacent child requires $z_\ell-1\neq z_{\ell-1}$.
\end{definition}



\begin{figure}
    \centering
    \includegraphics[width=0.75\linewidth]{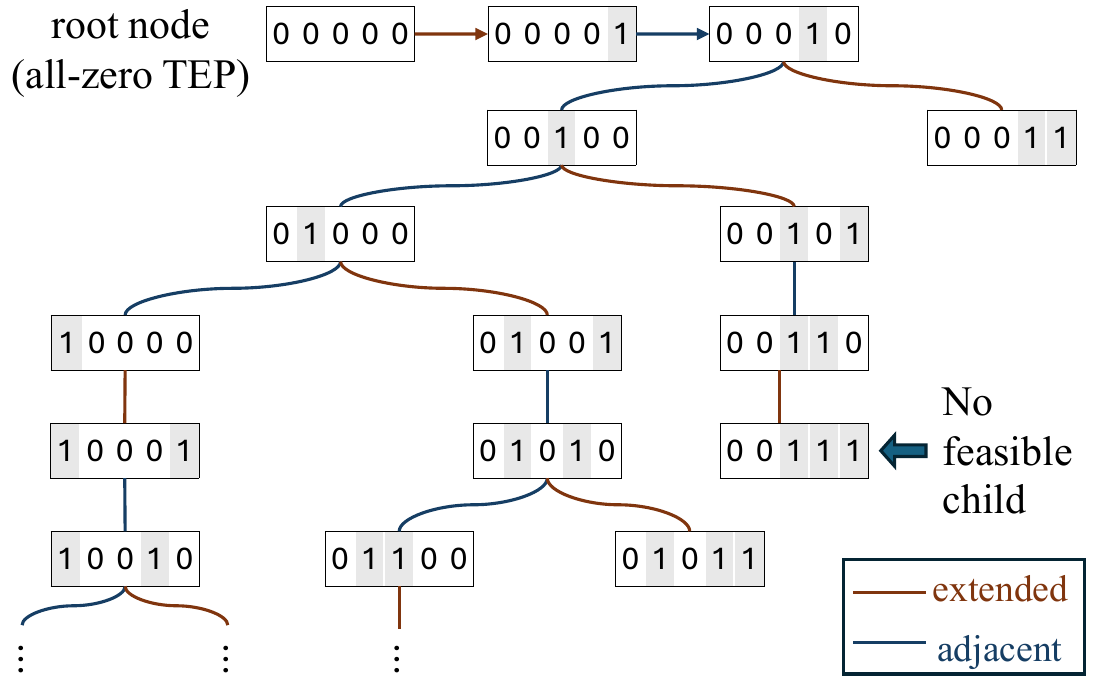}
    \vspace{-0.2cm}
    \caption{{The TEP tree for $k=5$ and $m=3$}.}
    \vspace{-0.2cm}
    \label{fig:placeholder2}
\end{figure}

According to Definition \ref{def::TEP}, a TEP tree is fully determined by message length $k$ and decoding order $m$, and it 
enumerates all TEPs with $w({\mathbf{e}})\le m$ exactly once. 
{Figure~\ref{fig:placeholder2} shows} the TEP tree for $k=5$ and  $m=3$.
For the root node $\mathcal{Z}_0=\varnothing$ {(i.e., 00000)},  only the extended child {\{5\} (i.e., 00001) exists}.
For the leaf TEP $\{{3},4,5\}$ {(i.e., 000111)},  neither  extended child nor  adjacent child exists. The tree contains exact 26 nodes, enumerating all valid TEPs with $w(\mathbf{e})\leq 3$. Note that from the root node, the weight-3 TEP 00111 can be reached with 6 steps, compared to $\sum_{i=0}^2\binom{5}{i} = 16$ steps in conventional OSD. 

To formulate MCTS search on this tree, we define: 
\begin{itemize}
    \item 
    {\textbf{State} $\mathbf{s}_t=\mathbf{e}_t$: The current TEP $\mathbf{e}_t$ represented by a tree node. Particularly, $\mathbf{s}_0=\mathbf{0}$ is the root all-zero TEP.}

    \item \textbf{Action}: 
    At $\mathbf{s}_t$, action $a_t\in \mathcal{A}=\{\texttt{extend}, \texttt{adjacent}\}$ transitions to the extended or adjacent child TEP.

    \item {\textbf{Reward}: Function $R(\mathbf{s}_t)$} assigns large positive rewards to nodes that can lead to the target TEP $\mathbf{e}^*$, while nodes that cannot reach $\mathbf{e}^*$ receive negative rewards. 
    \item {\textbf{Action value}: $Q(\mathbf{s}_t, a_t)$ denotes the maximum cumulative reward obtainable from state $\mathbf{s}_t$ by taking action $a_t$.}
\end{itemize}

\begin{remark}[Target TEP and codeword]
In practice, the target TEP $\mathbf{e}^*$ leading to the transmitted codeword is unknown during decoding. Therefore, for training, we obtain $\mathbf{e}^*$ corresponding to the MLD codeword, rather than the actual transmitted codeword. Our policy network learns to find MLD solutions under different noise realizations.
\end{remark}

\vspace{-0.3em}
\subsection{MCTS over TEP Tree} \label{sec::MCTS}
\vspace{-0.3em}

\begin{figure}[t]
    \centering
    \includegraphics[width=0.95\linewidth]{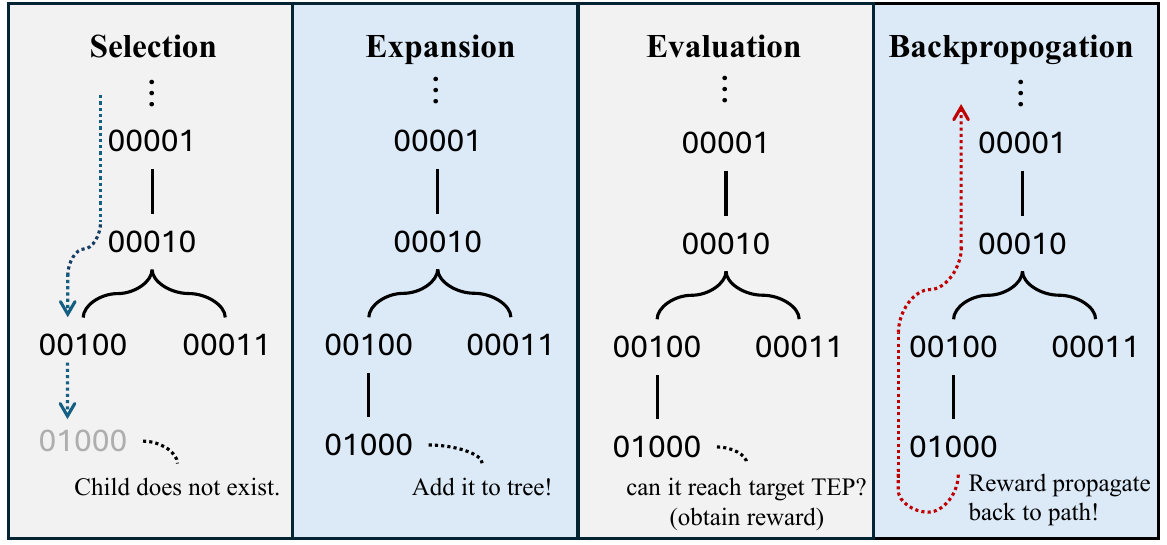}
    \vspace{-0.2cm}
    \caption{ An MCTS iteration on a TEP tree. MCTS reaches TEP 00100, selects an unexpanded child 01000, then expands, evaluates, and backpropagates.}
    \vspace{-0.cm}
    \label{fig:placeholder3}
\end{figure}

The MCTS search begins with a tree containing only the root node (all-zero TEP), and the tree is incrementally expanded as MCTS performs multiple searches.  Each search iteration follows the standard MCTS framework, i.e., steps including \textit{Selection}, \textit{Expansion}, \textit{Simulation} and \textit{Backpropagation}. Figure \ref{fig:placeholder3} illustrates an example. A policy neural network guides \textit{Selection}, and is iteratively trained on MCTS search statistics. The training will be explained in Section \ref{sec::training}.

\subsubsection{Selection}
Starting from the root state $\mathbf{s}_0$, we descend the tree by repeatedly selecting the child with the highest PUCT  (Predictor and Upper Confidence bounds \cite{lake2015human}) score, i.e.,
\begin{equation}
\label{eq:puct}
\mathrm{PUCT}(\mathbf{s}_t,a_t)= Q(\mathbf{s}_t,a_t) \!+\! c_{\mathrm{puct}}\,p(\mathbf{s}_t,a_t)\,\frac{\sqrt{N(\mathbf{s}_t)}}{1\!+\!N(\mathbf{s}_t,a_t)},
\end{equation}
where $p(\mathbf{s}_t,a_t)$ is the prior policy from a neural network, $N(\mathbf{s}_t)$ is the visit count for state $\mathbf{s}_t$ and $N(\mathbf{s}_t,a_t)$ is the number of times action $a_t$ has been selected from state $\mathbf{s}_t$. The hyperparameter $c_{\mathrm{puct}}$ controls the exploration-exploitation trade-off. 
At each state $\mathbf{s}_t$, MCTS selects action $$a_t^*=\arg\max_{a_t\in\mathcal{M}(\mathbf{s}_t)}\,\mathrm{PUCT}(\mathbf{s}_t,a_t),$$ where $\mathcal{M}(\mathbf{s}_t)$ is the legality mask for valid actions at state $\mathbf{s}_t$. If the resulting child $\mathbf{s}_{t+1} = f(\mathbf{s}_t, a_t^*)$ exists in the tree, descend to it and repeat. Otherwise, proceed to \textit{Expansion}.

\subsubsection{Expansion}
At state $\mathbf{s}_t$, the action $a_t^*$ creates the new child node $\mathbf{s}_{t+1} = f(\mathbf{s}_t, a_t^*)$.  
We add $\mathbf{s}_{t+1}$ to the tree and initialize $N(\mathbf{s}_{t+1})=1$, $N(\mathbf{s}_t, a_t^*)=0$, and $Q(\mathbf{s}_t, a_t^*)=0$. We then proceed to \textit{Simulation}. 

\subsubsection{Simulation {(Evaluation)}}
Instead of performing simulations as in AlphaZero, we can directly evaluate whether each newly expanded child can reach the target TEP $\mathbf{e}^*$ 
using the tree structure. Let the new TEP $\mathcal{Z}_{\mathbf{e}_{t+1}} = \{z_1, \ldots , z_\ell\}$ and the target TEP $\mathcal{Z}_{\mathbf{e}^*} = \{z_1^*, \ldots, z_{\ell^*}^*\}$. We note that $\mathcal{Z}_{\mathbf{e}^*}$ can be reached from $\mathcal{Z}_{\mathbf{e}_{t+1}}$ if and only if
\circled{1} $\ell \leq \ell^*$, 
\circled{2} $z_i = z_i^*$ for $i \in \{1,2,\ldots, \ell-1\}$, 
and \circled{3} $z_\ell > z_\ell^*$. We omit the proof.

For each $\mathbf{s}_{t+1} = \mathbf{e}_{t+1}$, we then obtain the candidate codeword $\mathbf{c}_{\mathbf{e}_{t+1}} = ({\mathbf{b}}_0 \oplus \mathbf{e}_{t+1})\mathbf{G}$ and assign the reward
\begin{equation}
\label{equ::reward}
R(\mathbf{s}_{t+1}) = \begin{cases}
+100, & \text{if } \mathcal{Z}_{\mathbf{e}_{t+1}} \text{ can reach } \mathcal{Z}_{\mathbf{e}^*} \\
-d(\mathbf{c}_{\mathbf{e}_{t+1}},\mathbf{r}), & \text{otherwise,}
\end{cases}
\end{equation}
providing positive reward for paths toward the target TEP.

\subsubsection{Backpropagation} \label{sec::backpropogation}
We propagate the reward of newly expanded child back along the entire path. 
Let $\mathcal{T}=\{(\mathbf{s}_0,a_0^*),(\mathbf{s}_1,a_1^*),\ldots,(\mathbf{s}_t,a_t^*)\}$ be the trajectory leading to $\mathbf{s}_{t+1}$.
For each ancestor state-action pair  $(\mathbf{s}_i, a_i^*)$, $i \in\{0, \ldots, t\}$, update $N(\mathbf{s}_i) \leftarrow N(\mathbf{s}_i) + 1$, $N(\mathbf{s}_i, a_i^*) \leftarrow N(\mathbf{s}_i, a_i^*) + 1$, and $Q(\mathbf{s}_i, a_i^*) \leftarrow \max\{Q(\mathbf{s}_i, a_i^*), R(\mathbf{s}_{t+1})\}$.

After backpropagation, one search iteration completes.  The process repeats for a maximum of $K$ iterations or until the target TEP $\mathbf{e}^*$ is encountered.

\vspace{-0.3em}
\subsection{Training } \label{sec::training}
\vspace{-0.3em}

For training, a large collection of codeword transmissions is simulated over an AWGN channel to produce the received vectors $\mathbf{r}$ with LLR $\boldsymbol{\ell}$, as described in Section \ref{sec::model}. Each received vector is decoded by the original OSD with a sufficiently large order to obtain the near-MLD codeword $\mathbf{c}_{\mathbf{e}^*}$. The target TEP $\mathbf{e}^*$ is determined by the difference pattern between the first $k$ bits of $\mathbf{c}_{\mathbf{e}^*}$ and the hard-decision vector $\mathbf{b}_0$ derived from $\mathbf{r}$. Let $\mathcal{D}_{\mathrm{train}} = \{(\mathbf{r}, \mathbf{c}_{\mathbf{e}^*,}
\mathbf{e}^*)\}$  denote that dataset.

The neural network policy used in 
\eqref{eq:puct} is defined as
\begin{equation}
\label{eq:policy_nn}
p(\mathbf{s}_t,a_t) = f_\theta\big(\mathbf{e}_t,\mathbf{c}_{\mathbf{e}_t},d(\mathbf{c}_{\mathbf{e}_t},\mathbf{r}), \mathbf{G}_{\text{flat}}, \bar{\boldsymbol{\ell}}\big),
\end{equation}
  where $d(\mathbf{c}_{\mathbf{e}_t},\mathbf{r})$ is the Euclidean distance,$\mathbf{G}_{\text{flat}} \in \mathbb\{0,1\}^{n\times k}$ is the generator matrix reshaped into a vector by concatenating its rows, and $\bar{\boldsymbol{\ell}} = (\boldsymbol{\ell} - \mu_{\ell})/\sigma_{\ell}$ is the normalized LLR with mean $\mu_{\ell}$ and standard deviation $\sigma_{\ell}$ computed from $\boldsymbol{\ell}$. The network outputs action probabilities for $a_t \in \mathcal{M}(\mathbf{s}_t)$.

For each training sample $(\mathbf{r}, \mathbf{c}_{\mathbf{e}^*,}
\mathbf{e}^*)\in \mathcal{D}_{\mathrm{train}} $, we perform $K$ independent MCTS episodes, each starting from the root state $\mathbf{s}_0 = \mathbf{0}$  following the four-step procedure in in Section~\ref{sec::MCTS}. Each episode terminates when the target TEP is found ($\mathbf{e}_t = \mathbf{e}^*$), a step budget $M$ is exhausted, or no legal actions remain. After the $K$ episodes, 
we compute the visit-count distribution ${\pi}_t(a_t) = N(\mathbf{s}_t, a_t) / \sum_{a \in M(\mathbf{s}_t)} N(\mathbf{s}_t, a)$ as the training target for each explored state $\mathbf{s}_t$ . The state-target pairs $(\mathbf{s}_t, {\pi}_t)$ are collected in replay buffer $\mathcal{B}$. When $|\mathcal{B}| = B$, we minimize the cross-entropy loss
\begin{equation}
\label{eq:policy_loss}
\mathcal{L}_{\text{pol}} = -\frac{1}{B}\sum_{(\mathbf{s}_t,{\pi}_t)\in\mathcal{B}} \sum_{a_t\in M(\mathbf{s}_t)} {\pi}_t(a_t)\log p(\mathbf{s}_t,a_t)
\end{equation}
via stochastic gradient descent (SGD) with learning rate $\eta$ and Adam optimizer. Then, we clear the buffer. This training procedure follows AlphaZero~\cite{silver2017mastering}, where MCTS refines the visit-count distributions for the neural network to learn. The training process of a single epoch is summarized in Algorithm~\ref{alg:mcts_train}.


\begin{algorithm}[h]
\small
\caption{Training policy network via MCTS}
\label{alg:mcts_train}
\DontPrintSemicolon
\SetInd{0.2em}{0.6em}
\SetAlgoVlined
\SetKwInput{KwInput}{Input}
\SetKwInput{KwOutput}{Output}

\KwInput{Dataset $\mathcal{D}_{\text{train}}$, Generator matrix $\mathbf{G}$, max steps $M$, MCTS episodes $K$, batch size $B$, decoding order $m$}
\KwOutput{Well-trained policy $f_\theta(\cdot)$}

Initialize replay buffer $\mathcal{B}\leftarrow\emptyset$.\;

\ForEach{$(\mathbf{r}, \mathbf{c}_{\mathbf{e}^*,}
\mathbf{e}^*)\in\mathcal{D}_{\text{train}}$}{
  Initialize $q\leftarrow 0$ and $\mathbf{b_0}$ (hard-decision of first $k$ bits of $\mathbf{r}$).\;
  \For{$q = 1,\ldots,K$}{
    Initialize the root node of TEP tree $\mathbf{s_0=\mathbf{e}_0=\mathbf{0}}$.\;
    \For{$t = 0,\ldots,M$}{
      For state $\mathbf{s}_t = \mathbf{e}_t$, obtain $\mathbf{c}_{\mathbf{e}_{t}} \!\!=\!\! ({\mathbf{b}}_0 \!\oplus\! \mathbf{e}_{t})\mathbf{G}$.\;
      \textbf{if} {$\mathbf{e}_t=\mathbf{e}^\ast$} \textbf{then break}\\
      Select action $a_t^*=\arg\max_{a_t\in\mathcal{M}(\mathbf{s}_t)}\,\mathrm{PUCT}(\mathbf{s}_t,a_t)$.\;
      \textbf{if} \textit{no legal action $a_t^*$} \textbf{then break}\\
      Transit to child state $\mathbf{s}_{t+1}$ by applying $a_t^*$.\\
        \uIf{The child $\mathbf{s}_{t+1}$ exists}{\textbf{continue}}
         \Else{
            Add $\mathbf{s}_{t+1}$ to the TEP tree.\\
            Initialize $N(\mathbf{s}_{t+1})=1$, $N(\mathbf{s}_t,a_t^\ast){=}0$, $Q(\mathbf{s}_t,a_t^\ast){=}0$.\\
            
            Obtain $R(\mathbf{s}_{t+1})$ according to \eqref{equ::reward}.\\
            Backpropagation as detailed in Section \ref{sec::backpropogation}.
          }
    }
  }
  Obtain ${\pi}_t(a_t)$ for all visited states, and $(\mathbf{s}_t,\pi_t)$ to $\mathcal{B}$. \\
  \If{$|\mathcal{B}| = B$}{
    Calculate $\mathcal{L}_{\text{pol}}$ and optimize the policy. Set $\mathcal{B} \leftarrow \varnothing$.
  }
}
\end{algorithm}

\vspace{-0.3em}
\subsection{{Inference (Online Decoding)}}\label{inference}
\vspace{-0.3em}
During inference, unlike training where the tree is expanded incrementally, the TEP tree is fully pre-generated and stored. The trained policy network guides a depth-first search over the TEP tree without MCTS statistics. Starting from the root state $\mathbf{s}_0$, we iteratively select the action according to:
\begin{equation}\label{a_infer}
a_t^* = \arg\max_{a_t\in M(\mathbf{s}_t)} p(\mathbf{s}_t,a_t),
\end{equation}
where $p(\mathbf{s}_t,a_t)$ is the output from trained policy $f_\theta$ in \eqref{eq:policy_nn}. 

The MLD results $\mathbf{c_{\mathbf{e^*}}}$ is unknown and must be searched by the decoder. At each visited state $\mathbf{s}_t$, we generate the candidate codeword $\mathbf{c}_{\mathbf{e}_t} = (\hat{\mathbf{b}}_0 \oplus \mathbf{e}_t)\mathbf{G}$ and evaluate it against a stopping criterion. In this paper, we consider either of two criteria:
\begin{itemize}
    \item \emph{Practical Probability-based Stopping:} {
We use the stopping condition from PB-OSD~\cite[Eq. (13)-(14)]{yue2022ordered}, which computes the success probability $P_{\mathrm{s}}(\mathbf{c}_{\mathbf{e}_t} \mid \mathbf{d}_t)$ for each codeword $\mathbf{c}_{\mathbf{e}_t}$, where $\mathbf{d}_t$ is the difference pattern between $\mathbf{c}_{\mathbf{e}_t}$ and the hard decision of $\mathbf{r}$. The decoding terminates if $P_{s}(\mathbf{c}_{\mathbf{e}_t} \mid \mathbf{d}_t) \geq \tau$ for a threshold $\tau \in [0,1]$, and outputs $\hat{\mathbf{c}} = \mathbf{c}_{\mathbf{e}_t}$ as the decoded result. We apply $\tau = 0.9$.
}
    \item \emph{Perfect Stopping:} {The search terminates immediately upon finding the MLD codeword $\mathbf{c_{\mathbf{e^*}}}$}, which is then output as $\hat{\mathbf{c}}$. This criterion is impractical and only used to evaluate the search efficiency to reach the MLD solution.
\end{itemize}
We note that the proposed policy-guided search can be used with any other stopping rule from the literature, e.g., \cite{jin2006probabilistic}.

\begin{figure}[t]
    \centering
    \includegraphics[width=0.75\linewidth]{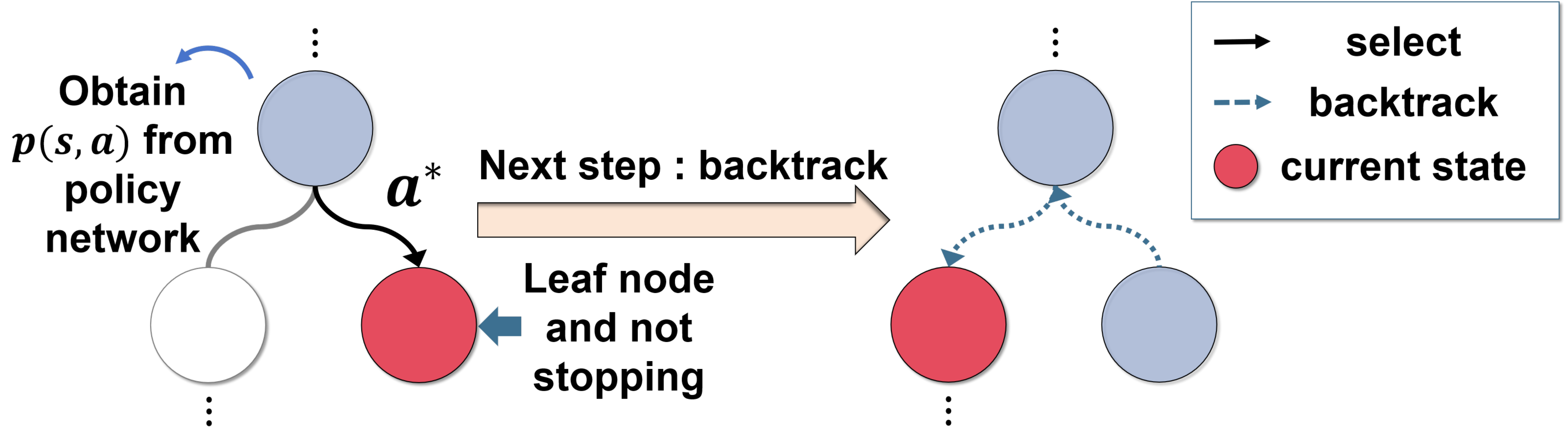}
    \caption{Example of depth-first TEP search. The search first visits the left child via action \(a^*\), then backtracks to explore the right child.}
    \label{fig:DFS}
    \vspace{-0.2em}
\end{figure}

When the stopping criterion is not satisfied at a leaf node, we backtrack to the nearest ancestor with unexplored children and continue with its next highest-probability action (see Fig.~\ref{fig:DFS}). If no stopping criterion is met before the step budget $M$ is exhausted, we output the candidate with minimum Euclidean distance, i.e., $\hat{\mathbf{c}} = \arg\min_{\mathbf{c} \in \mathcal{C}_{\text{visited}}} d(\mathbf{c,\mathbf{r}})$, where $\mathcal{C}_{\text{visited}}$ is the set of all visited TEPs' associated codewords. The inference is summarized in Algorithm~\ref{alg:mcts_infer}.

\begin{remark}
    The depth-first strategy with backtracking ensures complete coverage of TEPs up to order $m$ to guarantee BLER if needed, while prioritizing high-probability paths first.
    
\end{remark}

\begin{algorithm}[h]
\small
\caption{Inference: {Policy}-Guided Decoder}
\label{alg:mcts_infer}
\DontPrintSemicolon
\SetInd{0.2em}{0.6em}
\SetAlgoVlined
\SetKwInput{KwInput}{Input}
\SetKwInput{KwOutput}{Output}

\KwInput{Received signal $\mathbf{r}$, generator matrix $\mathbf{G}$, decoding order $m$, full TEP tree, maximum steps $M$}
\KwOutput{Decoded results  $\hat{\mathbf{c}}$}

Initialize $t\leftarrow 0$, $d_{\min} \leftarrow \infty$, and $\mathbf{b_0}$.\;

\For{$t = 0,\ldots,M$}{
  For state $\mathbf{s}_t = \mathbf{e}_t$, obtain codeword $\mathbf{c}_{\mathbf{e}_{t}} = ({\mathbf{b}}_0 \oplus \mathbf{e}_{t})\mathbf{G}$.\;
  \uIf{$d(\mathbf{c}_{\mathbf{e}_{t}},\mathbf{r}) < d_{\min}$}{
    $d_{\min}\leftarrow d(\mathbf{c}_{\mathbf{e}_{t}},\mathbf{r})$ and $\hat{\mathbf{c}} = \mathbf{c}_{\mathbf{e}_{t}}$.\;}
  \textbf{if} $\mathbf{c}_{\mathbf{e}_{t}}$ satisfies stopping criterion \textbf{then return} $\hat{\mathbf{c}}$\\
  Select action $a_t^*$ according to~\eqref{a_infer}.\;
  \uIf{No legal action $a_t^*$}{
    \textbf{Backtrack} to most recent ancestor with an unexplored child; set $\mathbf{s}_{t+1}$ to that child.\;
  }\Else{
    Transit to child state  $\mathbf{s}_{t+1}$ by applying $a_t^*$.\;
  }
}
\textbf{return} $\hat{\mathbf{c}}$\;
\end{algorithm}

\vspace{-0.3em}
\subsection{Complexity {Consideration}}
\vspace{-0.3em}
We analyze the decoding complexity of the proposed MCTS decoder and compare it with OSD variants. 

\subsubsection{Neural network complexity.}
We intentionally design a lightweight policy network with $h=3\sim 6$ hidden layers and 128 hidden units per layer. 
For input dimension $d_{\text{in}} = k + kn + 2n+1$ and two output actions according to \eqref{eq:policy_nn}, one forward pass costs $
C_{\text{nn}} = d_{\text{in}} \cdot 128 + (h-1) \cdot 128^2 + 128 \cdot 2 = \mathcal{O}(kn)$ FLOPs,
dominated by the $k\times n$ generator matrix. Since most tree nodes have only one legal action, 
for a tree traversal of $N_{\text{MCTS}}$ steps, the number of network calls is approximately $N_{\text{nn}}\approx N_{\text{MCTS}}/3$. 
For example, Fig.~\ref{fig:placeholder2} shows that reaching TEP 00111 requires 7 steps but only 2 network calls.

\subsubsection{TEP search complexity} 
Both order-$m$ OSD and non-GE OSD require to search $\sum_{j=0}^{m}\binom{k}{j}$ TEPs in the worst case. In contrast, in the proposed MCTS decoder, a well-trained policy network is able to select the shortest path to the target TEP. The worst-case complexity is therefore the maximum tree depth $m(2k-m+1)/2$, which is obtained by repeatedly selecting the adjacent action whenever legal. For instance, with $k=16$ and $m=3$, OSD requires $697$ TEPs, whereas MCTS examines at most $3\times(32-3+1)/2 = 45$ nodes.

\subsubsection{Overall complexity}

Let $N_{\text{non-GE}}$, $N_{\text{OSD}}$, and $N_{\text{MCTS}}$ denote the average number of TEP evaluations for non-GE OSD, standard OSD, and the proposed MCTS decoder, respectively. Each TEP requires one encoding operation $\mathbf{c}_{\mathbf{e}} = (\hat{\mathbf{b}}_0 \oplus \mathbf{e})\mathbf{G}$ at cost $\mathcal{O}(k(n-k))$. The overall decoding complexities are
\begin{align}
C_{\text{NonGE-OSD}} &=  \mathcal{O}(N_{\text{non-GE}} \cdot k(n-k)),  \notag\\
C_{\text{OSD}} &= \mathcal{O}(N_{\text{OSD}} \cdot k(n-k) + C_{\text{GE}}), \notag\\
C_{\text{MCTS}} &= \mathcal{O}(N_{\text{MCTS}} \cdot k(n-k) + (N_{\text{MCTS}}/3) \cdot kn), \notag
\end{align}
where $C_{\text{GE}}=n\cdot\min(k,n-k)^2$ represents the GE overhead in OSD. The proposed decoder achieves $N_{\text{MCTS}} \ll N_{\text{non-GE}}$ and avoids GE entirely compared to OSD.

\begin{table}[h]
    \centering
    \caption{Hyper-parameters of training}
    \begin{tabular}{|l|c|c|}
    \hline
    \textbf{Parameters} & \textbf{(32,16) eBCH code} & \textbf{(48,24) QR code} \\ \hline
    order $m$ & $5$ & $6$ \\ \hline
    Dataset size $|\mathcal{D}_{\mathrm{train}}|$ & $1.1\times10^5$ & $1.2\times10^6$ \\ \hline
    Maximum steps $M$ & $70$ & $129$ \\ \hline
    MCTS episodes $K$& $3000$ & $10000$ \\ \hline
    Hidden layers $h$& $3$ & $6$ \\ \hline
    Hidden units & \multicolumn{2}{c|}{$128$} \\ \hline
    Batch size $\mathcal{B}$ & \multicolumn{2}{c|}{$4096$} \\ \hline
    Learning rate $\eta$ & \multicolumn{2}{c|}{$10^{-4}$} \\ \hline
    Epochs & \multicolumn{2}{c|}{$150$} \\ \hline
     $c_{\text{puct}}$ in \eqref{eq:puct} & \multicolumn{2}{c|}{$1.38$} \\ \hline
    \end{tabular}
    \label{tab:hyperparameters}
    \vspace{-0.75em}
\end{table}

\vspace{-0.5em}
\section{Experiments and Results} \label{sec:results}
\vspace{-0.3em}
\subsection{Experimental Setup}
\vspace{-0.3em}
We evaluate the proposed decoding framework on both a $(32,16)$ extended BCH (eBCH) code and a $(48,24)$ quadratic residue (QR) code. For these codes, the non-GE OSD and the proposed MCTS decoder achieve near-MLD performance with order $m=5$ and $m=6$, respectively, while the standard OSD achieves it with order $m=3$. The non-GE OSD and OSD serve as benchmarks. All decoders use the same stopping criterion, allowing a fair comparison of their search efficiency in approaching near-MLD results.

 For training, each sample is obtained through simulation with an SNR uniformly sampled from the range $[0,5]$~dB. The samples are generated to cover as many distinct target TEPs as possible. The main hyper-parameters  are summarized in Table~\ref{tab:hyperparameters}. Note that maximum number $M$ of steps is set to the tree depth $m(2k-m+1)/2$ for efficient training.


During inference, $M$ is set to $\sum_{i=0}^m\binom{k}{i}$ to guarantee best BLER. Received signals are generated at each SNR, and policy-guided decoding is performed until 200 block errors are collected. All experiments are conducted on an Ubuntu 24.04 LTS server equipped with an Intel Xeon Platinum 8488C CPU (16 vCPUs, 32 GB RAM).

\begin{figure*}[t]
    \centering
    \begin{subfigure}{0.25\textwidth}
        \centering
        \includegraphics[width=\linewidth]{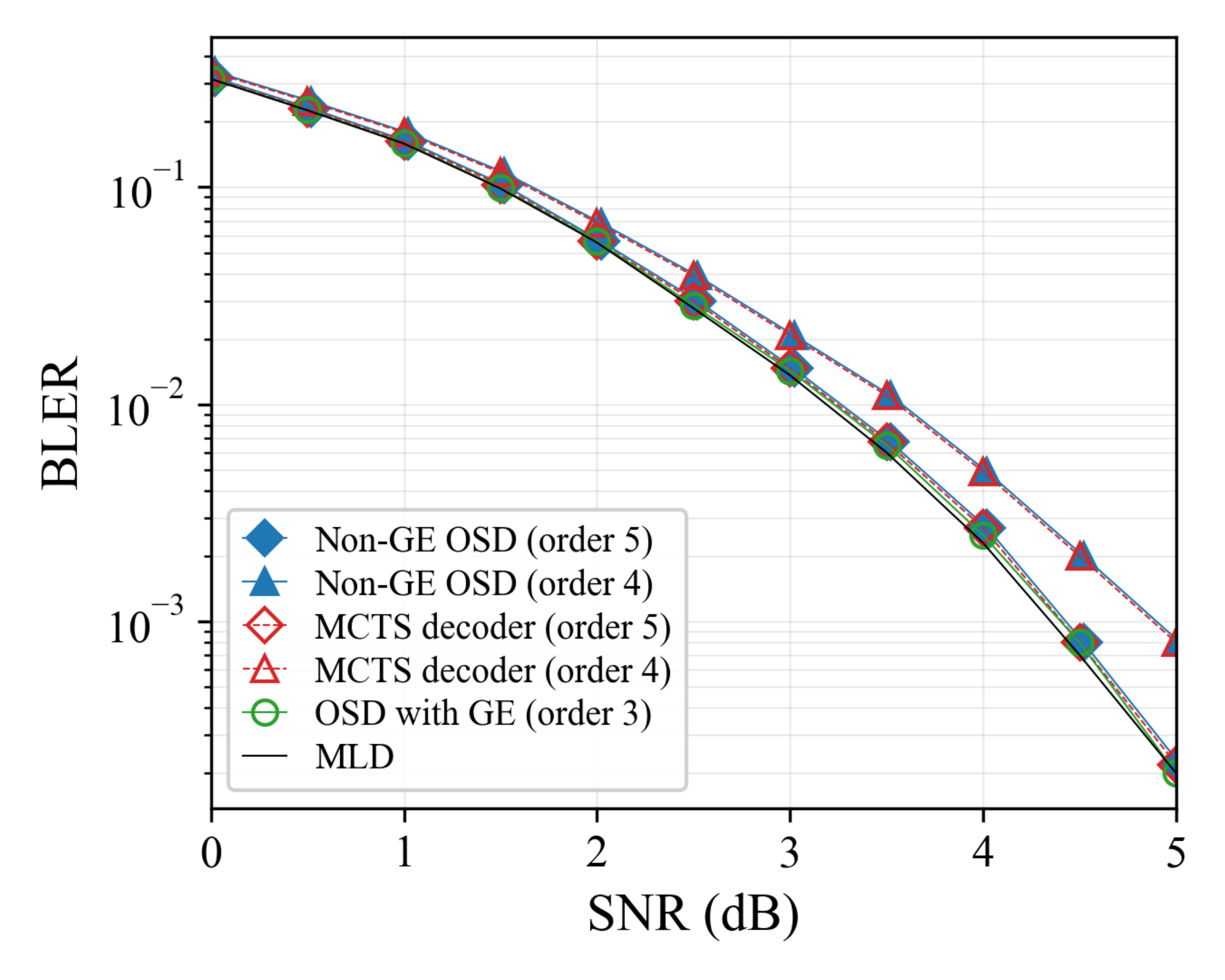}
                      \vspace{-2em}
        \subcaption{BLER}

        \label{fig:bler}
    \end{subfigure}\hfill
    \begin{subfigure}{0.25\textwidth}
        \centering
        \includegraphics[width=\linewidth]{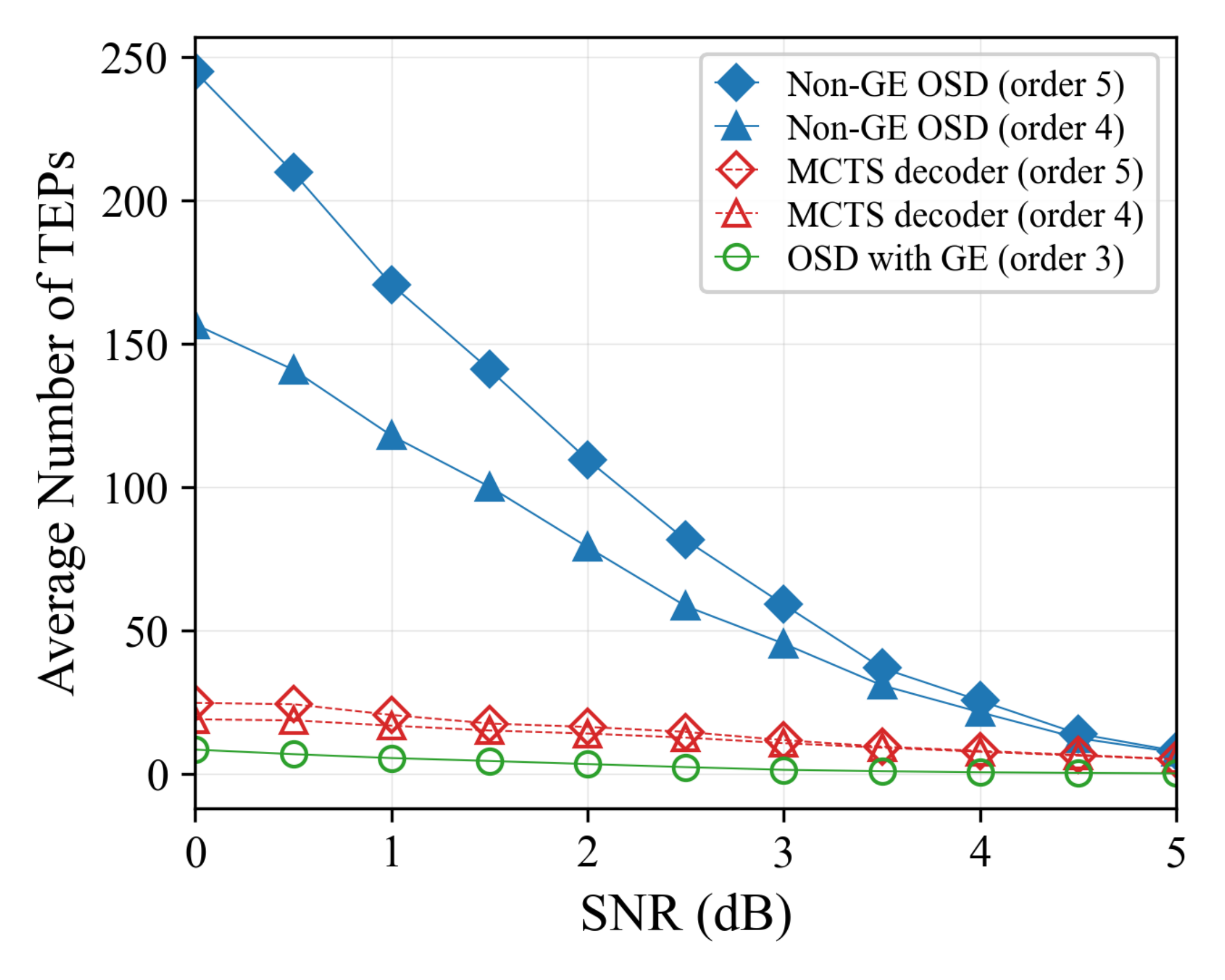}
                      \vspace{-2em}
        \subcaption{Avg. TEPs (perfect)}

        \label{fig:avgsteps-perfect}
    \end{subfigure}\hfill
    \begin{subfigure}{0.25\textwidth}
        \centering
        \includegraphics[width=\linewidth]{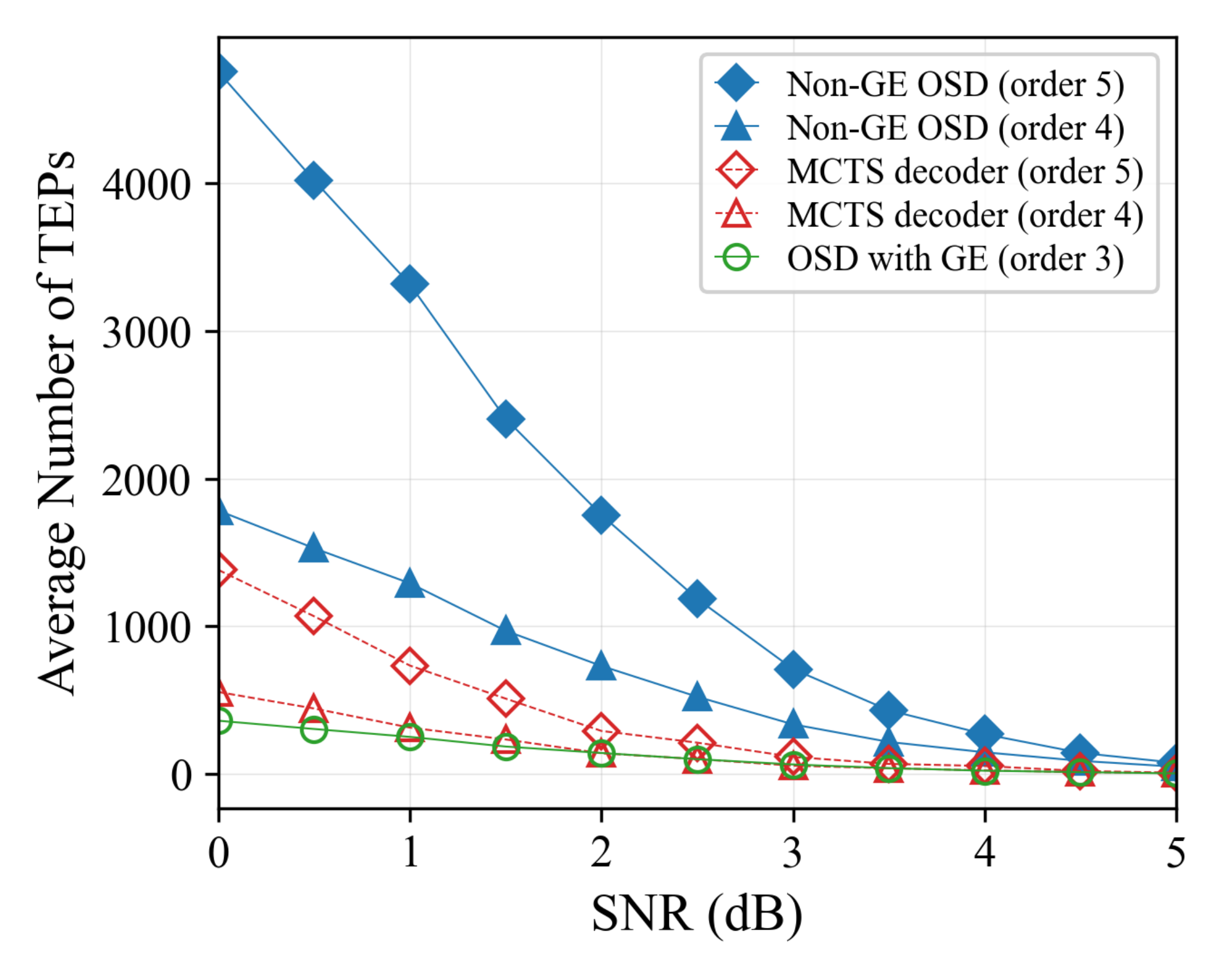}
                      \vspace{-2em}
        \subcaption{Avg. TEPs (practical)}
        \label{fig:avgsteps-prob}
    \end{subfigure}\hfill
    \begin{subfigure}{0.25\textwidth}
        \centering
        \includegraphics[width=\linewidth]{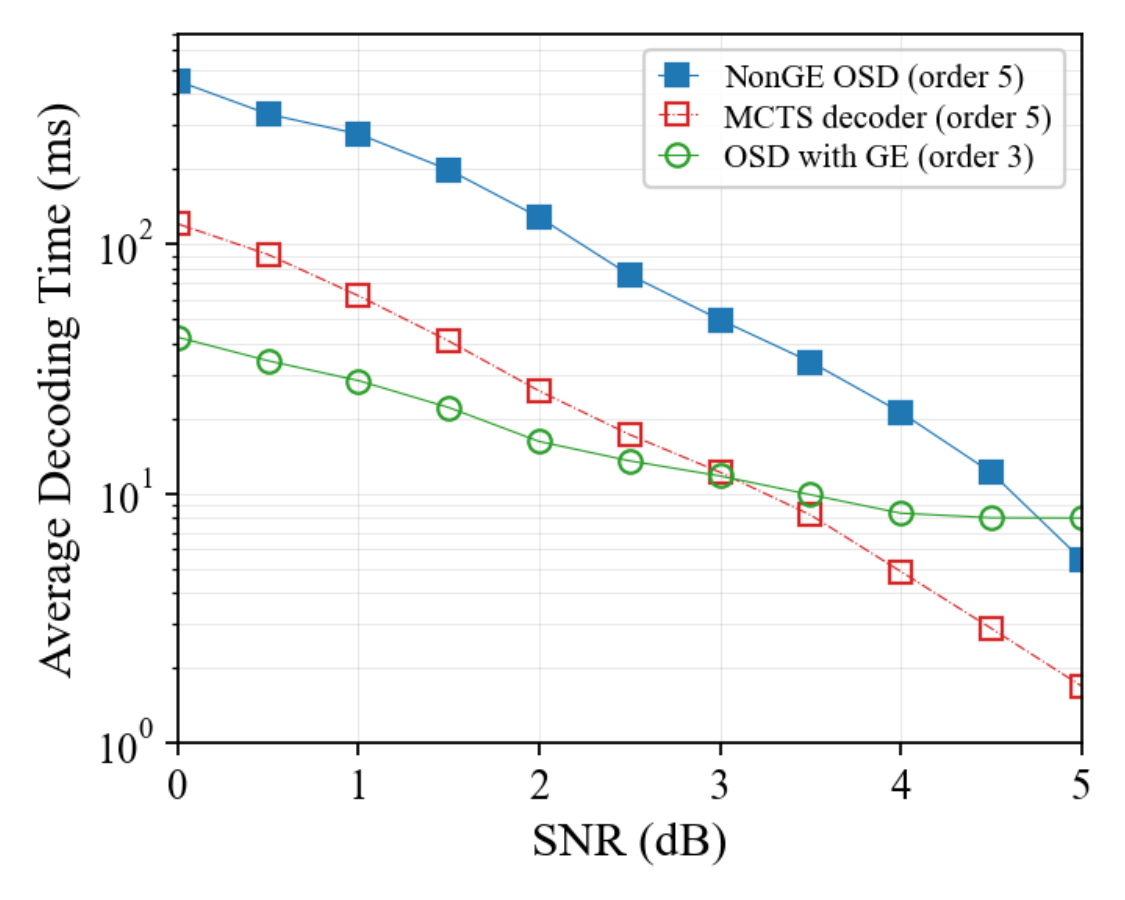}
                      \vspace{-2em}
        \subcaption{Decoding time (practical)}
        \label{fig:dctime}
    \end{subfigure}
          \vspace{-1.5em}
    \caption{Comparison of OSD, non-GE OSD and MCTS decoding for (32,16) eBCH code: (a) BLER performance; (b) average searched TEPs under perfect stopping criterion; (c) average searched TEPs under practical stopping criterion; (d) average decoding time under practical stopping criterion.}
    \label{fig:one-line-4}
    \vspace{-0.8em}
\end{figure*}

\begin{figure*}[t]
  \centering
    \begin{subfigure}{0.25\textwidth}
    \centering
    \includegraphics[width=\linewidth]{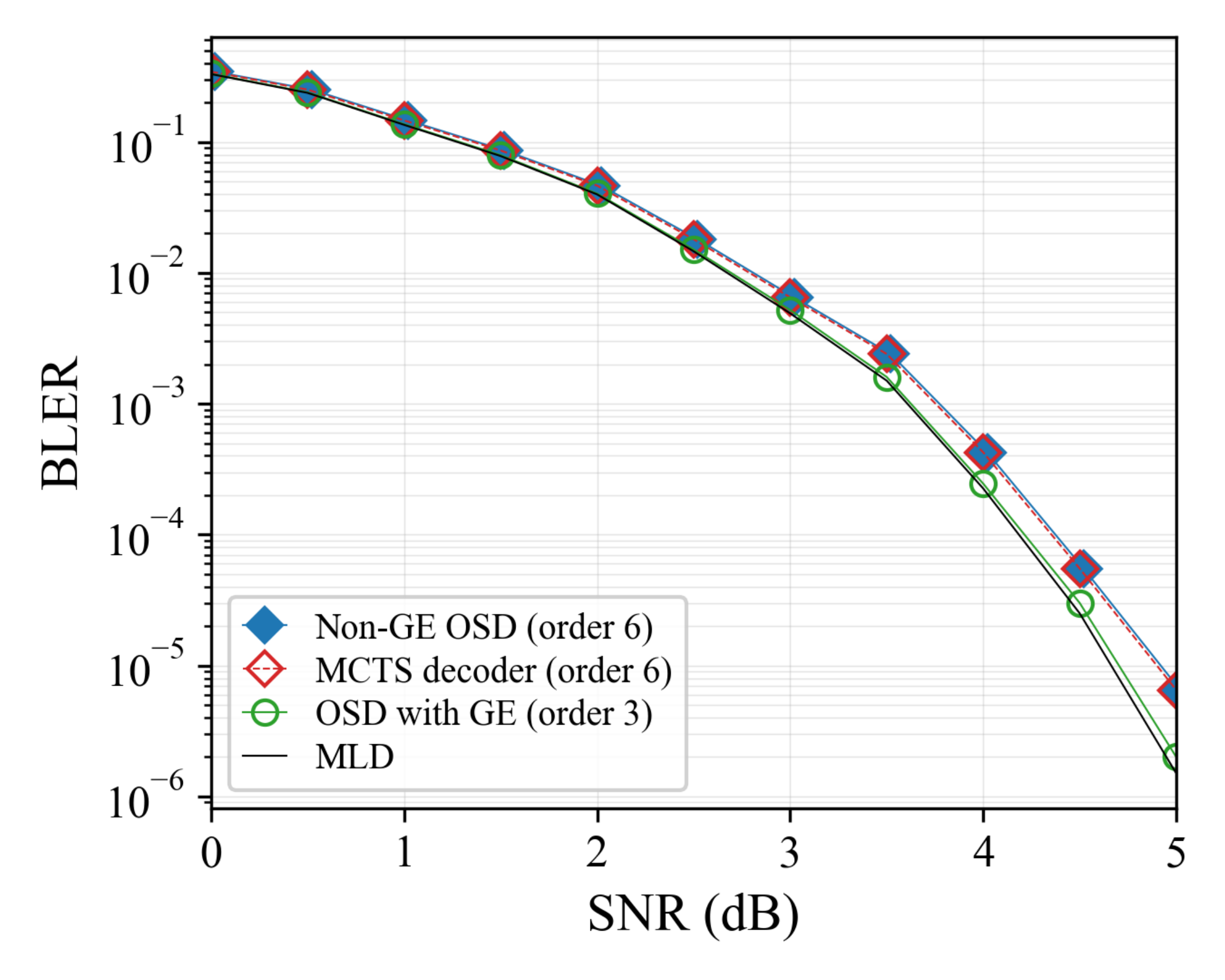}
    \vspace{-2em}
    \caption{BLER}
    \label{fig:bler4824}
  \end{subfigure}\hfill
  \begin{subfigure}{0.25\textwidth}
    \centering
    \includegraphics[width=\linewidth]{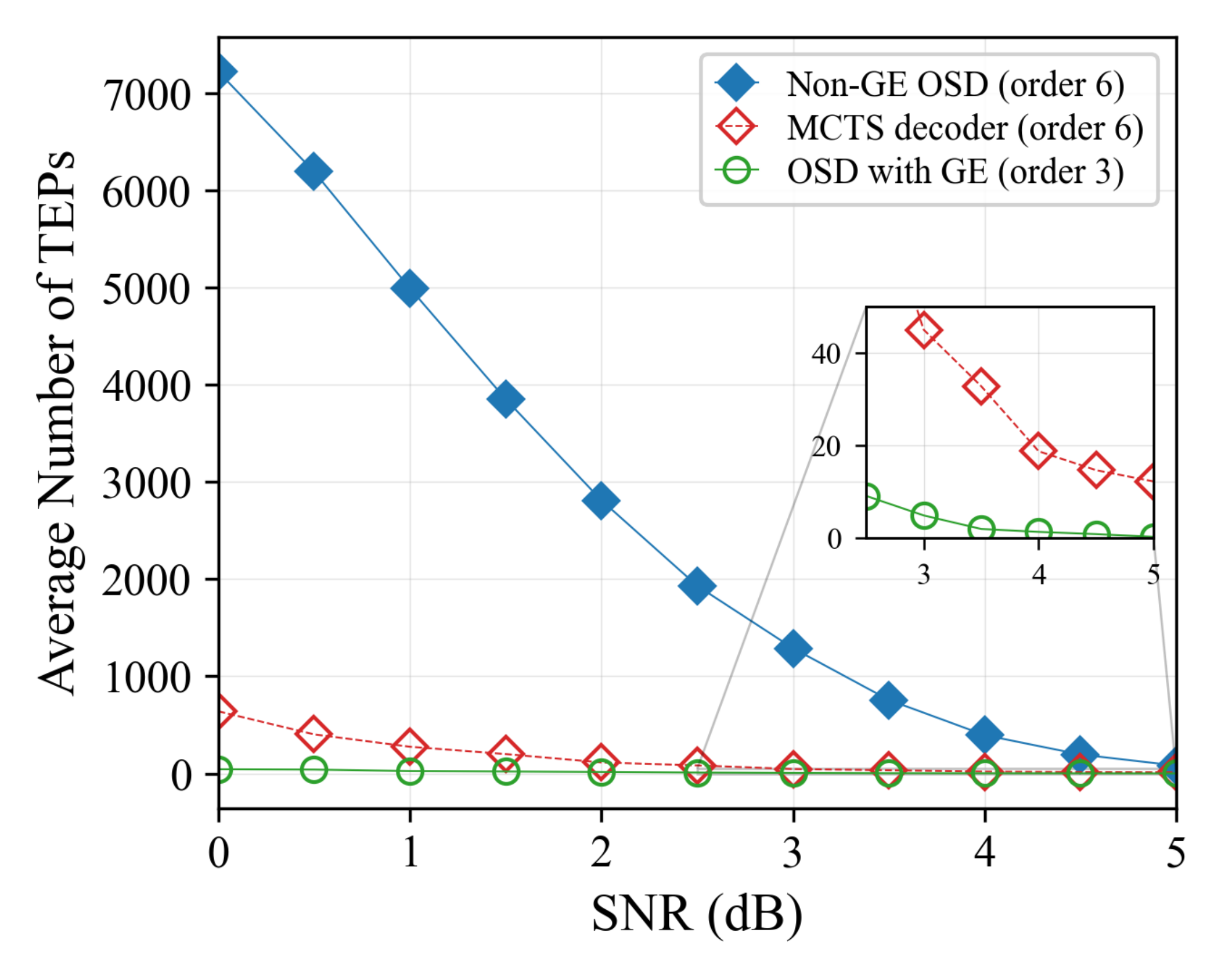}
    \vspace{-2em}
    \caption{Avg. TEPs (perfect)}
    \label{fig:avg_steps4824}
  \end{subfigure}\hfill
  \begin{subfigure}{0.25\textwidth}
      \centering
      \includegraphics[width=\linewidth]{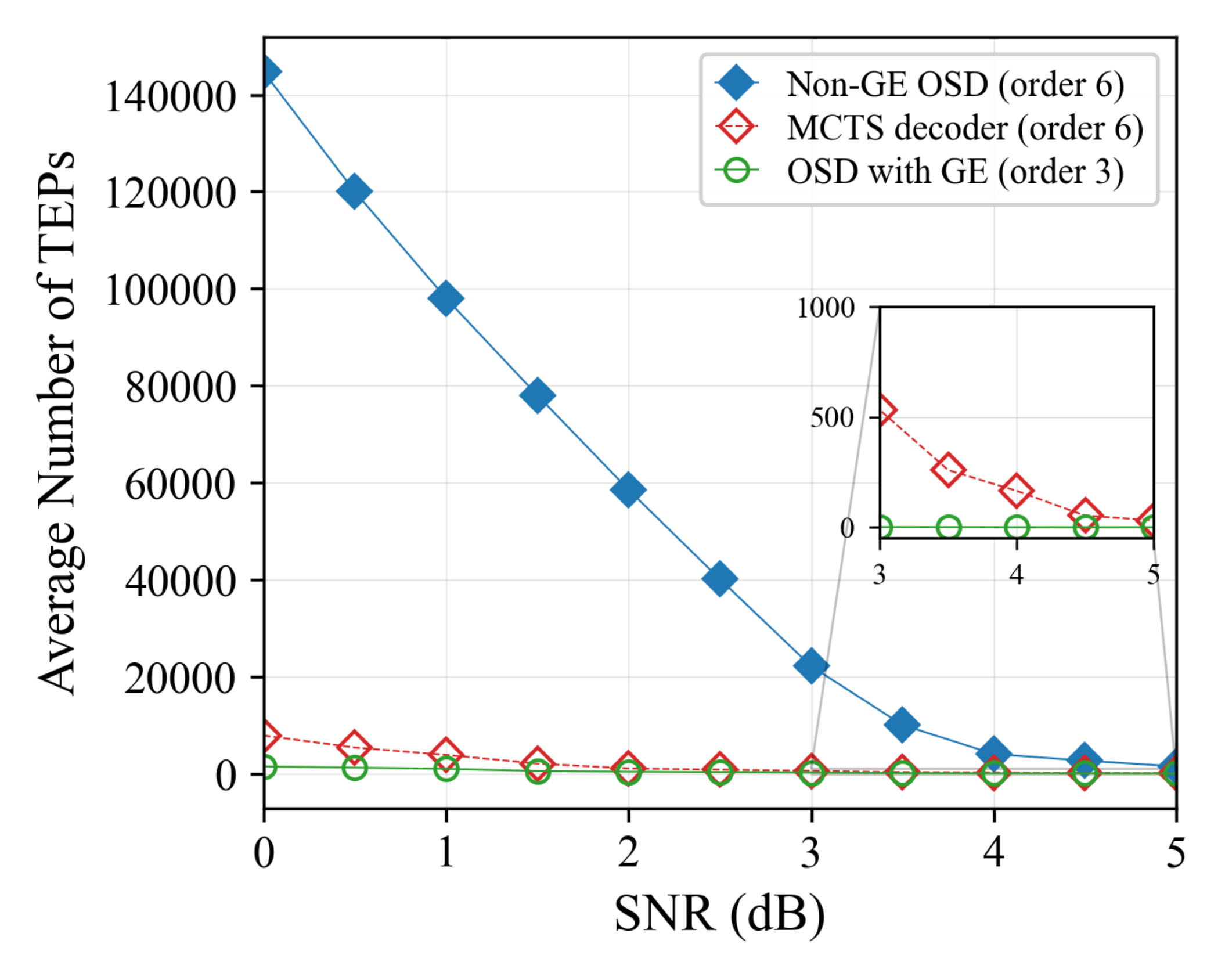}
      \vspace{-2em}
      \caption{Avg. TEPs (practical)}
      \label{fig:avg_steps4824_cs}
  \end{subfigure}\hfill
  \begin{subfigure}{0.25\textwidth}
      \centering
      \includegraphics[width=\linewidth]{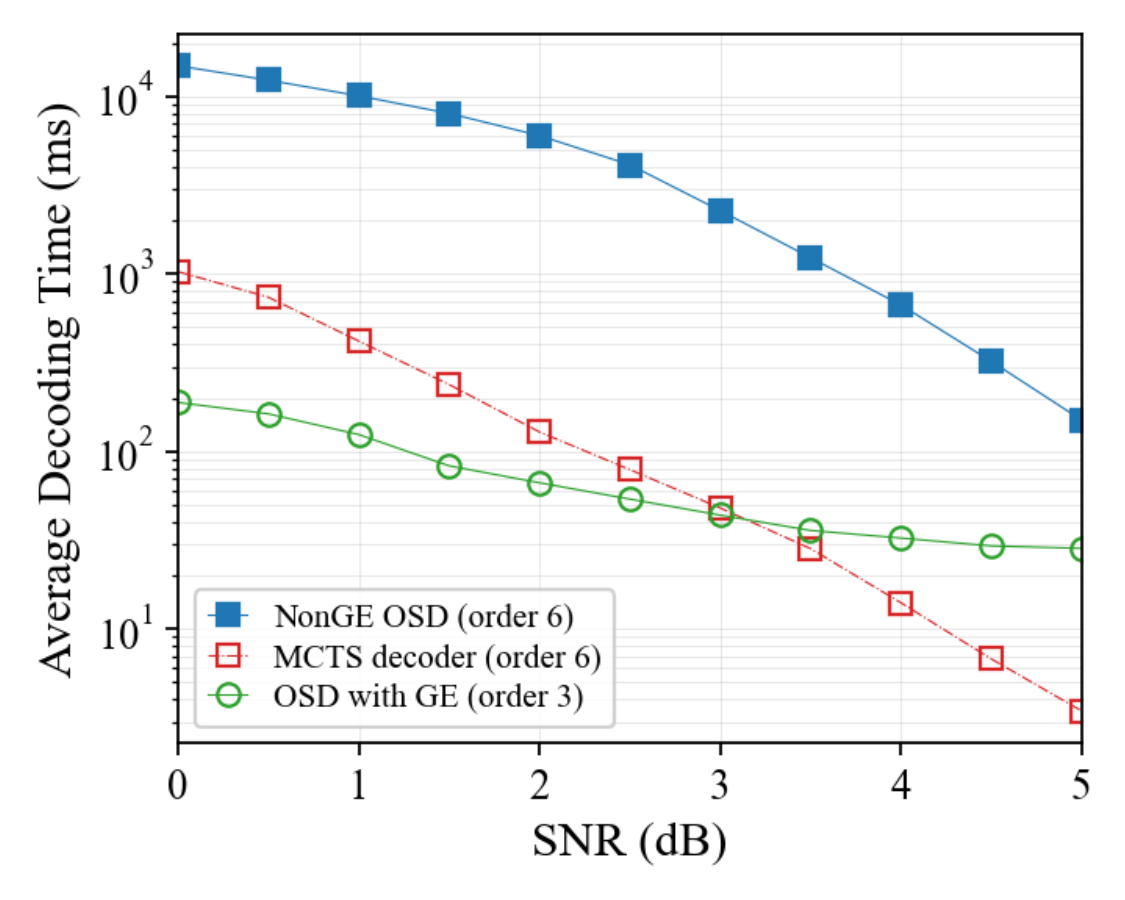}
      \vspace{-2em}
      \caption{Decoding time (practical)}
      \label{fig:dctime4824}
  \end{subfigure}
    \vspace{-1.5em}
  \caption{Comparison of OSD, non-GE OSD and MCTS decoding for (48,24) QR code over AWGN: (a) BLER performance; (b) average searched TEPs under perfect stopping criterion;(c)average searched TEPs under practical stopping criterion; (d) average decoding time under practical stopping criterion.}
  \label{fig:teps_bler_combo}
  \vspace{-0.8em}
\end{figure*}

\vspace{-0.3em}
\subsection{Simulation Results}
\vspace{-0.3em}
\subsubsection{$(32,16)$ eBCH code}
As shown in Figure~\ref{fig:bler}, for the $(32,16)$ eBCH code, the proposed decoder achieves near-MLD BLER performance with $m{=}5$. Lowering the order $m$ leads to a gradual performance degradation. It achieves the same BLER as non-GE OSD at the same order.

Figure~\ref{fig:avgsteps-perfect} illustrates the average number of TEPs searched under the perfect stopping criterion. Across all orders, the proposed MCTS decoder visits significantly fewer TEPs than the non-GE OSD. 
For example, at $m=5$ and SNR 0 dB, non-GE OSD visits over 250 TEPs, while the proposed decoder requires fewer than 20. We highlight that although the TEP count in non-GE OSD increases rapidly with $m$ due to exhaustive enumeration, the growth in MCTS decoding remains moderate.
Under the practical probability-based stopping criterion, as shown in Figure~\ref{fig:avgsteps-prob}, the proposed method still maintains high search efficiency (200 vs 100 vs 700 searches at 3 dB for MCTS, OSD, and non-GE OSD, respectively). It has only a slightly higher count than OSD but without the GE overhead.

Figure~\ref{fig:dctime} compares the decoding latency of OSD, non-GE OSD, and the proposed MCTS decoding, under the practical stopping. At high SNRs, the runtime of OSD remains nearly constant at around 10 ms, mainly due to the Gaussian elimination step. By removing this step, non-GE OSD achieves a shorter decoding time of about 5 ms.
At low SNRs, the MCTS decoder takes longer than OSD due to more search steps, yet it still outperforms OSD at much lower SNRs compared to non-GE OSD.

\subsubsection{$(48,24)$ QR code}
As shown in Figure~\ref{fig:bler4824}, for the $(48,24)$ QR code, the proposed MCTS decoding achieves near-MLD BLER performance with $m=6$. Figure~\ref{fig:avg_steps4824} and ~\ref{fig:avg_steps4824_cs} present the average number of visited TEPs under perfect and practical stopping criterion, respectively. The MCTS decoder requires a similar number of searches as OSD, both dramatically fewer than non-GE OSD (e.g., at 5 dB SNR,  31 for MCTS, 10 for OSD, and 1460 for non-GE OSD with practical stopping).


Fig.~\ref{fig:dctime4824} shows that MCTS decoding is clearly more  time-efficient for $(48,24)$ QR code, with practical stopping criterion. For example, at 5 dB SNR, the average decoding time is 3.5 ms for MCTS, 28 ms for OSD, and 150 ms for non-GE OSD. The delay of OSD saturates at high SNR due to GE overhead, whereas non-GE OSD suffers from excessive search steps leading to higher delays. The decoding time and TEP search counts are summarized in Table~\ref{tab:dec-time-steps}.


\begin{table}[t]
\centering
\caption{Decoding time (ms) and average searched TEPs for (48,24) QR code, under the practical stopping criterion.}
\vspace{-0.6em}
\label{tab:dec-time-steps}
\setlength{\tabcolsep}{3pt}       
\renewcommand{\arraystretch}{1.05}
\scriptsize 
\begin{tabular}{l|cc|cc|cc}
\hline
& \multicolumn{2}{c|}{\textbf{1 dB}}
& \multicolumn{2}{c|}{\textbf{3 dB}}
& \multicolumn{2}{c}{\textbf{5 dB}}\\
\textbf{Method} & \textbf{TEP} & \textbf{Time} & \textbf{TEP} & \textbf{Time} & \textbf{TEP} & \textbf{Time} \\
\hline
OSD   &  954.0  &  124.4   &  187.0  &  43.7   &  10.9  &  28.4   \\
Non-GE OSD      &  98016.6  &  10070.2   &  22110.0  &  2271.6   & 1460.8    &  150.1  \\
MCTS  &  3825.7  &  418.4   &  532.0  &  48.0   &  31.2  &  3.5   \\
\hline
\end{tabular}
\vspace{-0.5em}
\end{table}

\vspace{-0.5em}
\section{Conclusion}\label{sec:: conclusion}
\vspace{-0.3em}
In this paper, we proposed 
policy-guided MCTS decoder for short 
codes. The proposed decoder attains near-MLD BLER performance while substantially reducing the number of TEPs visited 
compared to non-GE OSD. At high SNRs, it achieves lower decoding latency than both non-GE OSD earlier and standard OSD. 
The proposed framework 
can be seamlessly applied to any existing OSD variants.

\vspace{-0.8em}
\bibliographystyle{IEEEtran}

\bibliography{ref}

@ARTICLE{8831394,
  author={Chen, Jienan and Chen, Siyu and Qi, Yunlong and Fu, Shengli},
  journal={IEEE Trans. Signal Processing}, 
  title={Intelligent Massive {MIMO} Antenna Selection Using Monte Carlo Tree Search}, 
  year={2019},
  volume={67},
  number={20},
  pages={5380-5390},
  doi={10.1109/TSP.2019.2940128}}

@article{silver2016mastering,
  title={Mastering the game of {Go} with deep neural networks and tree search},
  author={Silver, David and others},
  journal={Nature},
  volume={529},
  number={7587},
  pages={484--489},
  year={2016},
  publisher={Nature Publishing Group}
}

@article{silver2017mastering,
  title={Mastering the game of {Go} without human knowledge},
  author={Silver, David and others},
  journal={Nature},
  volume={550},
  number={7676},
  pages={354--359},
  year={2017},
  publisher={Nature Publishing Group UK London}
}

@article{yue2021probability,
  title={Probability-based ordered-statistics decoding for short block codes},
  author={Yue, Chentao and Shirvanimoghaddam, Mahyar and Park, Giyoon and Park, Ok-Sun and Vucetic, Branka and Li, Yonghui},
  journal={IEEE Commun. Lett.},
  volume={25},
  number={6},
  pages={1791--1795},
  year={2021},
  publisher={IEEE}
}

@inproceedings{yue2022ordered,
  title={Ordered-statistics decoding with adaptive {Gaussian} elimination reduction for short codes},
  author={Yue, Chentao and Shirvanimoghaddam, Mahyar and Vucetic, Branka and Li, Yonghui},
  booktitle={IEEE GLOBECOM Workshops, GC Wkshps - Proc.},
  pages={492--497},
  year={2022},
  organization={IEEE}
}

@article{shirvanimoghaddam2018short,
  title={Short block-length codes for ultra-reliable low latency communications},
  author={Shirvanimoghaddam, Mahyar and others},
  journal={IEEE Commun. Mag.},
  volume={57},
  number={2},
  pages={130--137},
  year={2018},
  publisher={IEEE}
}

@article{nachmani2019hyper,
  title={Hyper-graph-network decoders for block codes},
  author={Nachmani, Eliya and Wolf, Lior},
  journal={Adv. Neural Inf. Process. Syst.},
  volume={32},
  year={2019}
}

@article{nachmani2022neural,
  title={Neural decoding with optimization of node activations},
  author={Nachmani, Eliya and Be’ery, Yair},
  journal={IEEE Commun. Lett.},
  volume={26},
  number={11},
  pages={2527--2531},
  year={2022},
  publisher={IEEE}
}

@article{larue2022neural,
  title={Neural belief propagation auto-encoder for linear block code design},
  author={Larue, Guillaume and Dufrene, Louis-Adrien and Lampin, Quentin and Ghauch, Hadi and Othman, Ghaya Rekaya-Ben},
  journal={IEEE Trans. Commun.},
  volume={70},
  number={11},
  pages={7250--7264},
  year={2022},
  publisher={IEEE}
}

@article{lake2015human,
  title={Human-level concept learning through probabilistic program induction},
  author={Lake, Brenden M and Salakhutdinov, Ruslan and Tenenbaum, Joshua B},
  journal={Science},
  volume={350},
  number={6266},
  pages={1332--1338},
  year={2015},
  publisher={American Association for the Advancement of Science}
}

@article{yue2023efficient,
  title={Efficient decoders for short block length codes in {6G URLLC}},
  author={Yue, Chentao and Miloslavskaya, Vera and Shirvanimoghaddam, Mahyar and Vucetic, Branka and Li, Yonghui},
  journal={IEEE Commun. Mag.},
  volume={61},
  number={4},
  pages={84--90},
  year={2023},
  publisher={IEEE}
}

@article{yue2025guesswork,
  title={The Guesswork of Ordered Statistics Decoding: Guesswork Complexity and Decoder Design},
  author={Yue, Chentao and She, Changyang and Vucetic, Branka and Li, Yonghui},
  journal={IEEE Trans. Inform. Theory},
  year={2025},
  publisher={IEEE}
}

@ARTICLE{Fossorier1995OSD, 
author={M. P. C. Fossorier and Shu Lin}, 
journal={IEEE Trans. Inform. Theory}, 
title={Soft-decision decoding of linear block codes based on ordered statistics}, 
year={1995}, 
volume={41}, 
number={5}, 
pages={1379-1396}, 
doi={10.1109/18.412683}, 
ISSN={0018-9448}, 
month={Sep},}

@article{yu2025ordered,
  title={Ordered Statistics Derivative Decoding for Affine-Invariant Codes without {Gaussian} Elimination},
  author={Yu, Shuyan and Zhang, Bin and Huang, Qin},
  journal={IEEE Trans. Commun.},
  year={2025},
  publisher={IEEE}
}

@article{fossorier2024modified,
  title={Modified {OSD} algorithm with reduced {Gaussian} elimination},
  author={Fossorier, Marc and Shakiba-Herfeh, Mahdi and Zhang, Huazi},
  journal={IEEE Commun. Lett.},
  volume={28},
  number={8},
  pages={1755--1759},
  year={2024},
  publisher={IEEE}
}

@article{chen20235g,
  title={{5G}-advanced toward {6G}: Past, present, and future},
  author={Chen, Wanshi and Lin, Xingqin and Lee, Juho and Toskala, Antti and Sun, Shu and Chiasserini, Carla Fabiana and Liu, Lingjia},
  journal={IEEE J. Select. Areas Commun.},
  volume={41},
  number={6},
  pages={1592--1619},
  year={2023},
  publisher={IEEE}
}

@article{dufrene2025learning,
  title={Learning Linear Block Codes with Gradient Quantization},
  author={Dufr{\`e}ne, Louis-Adrien and Lampin, Quentin and Larue, Guillaume},
  journal={arXiv.2503.16169},
  year={2025}
}

@article{chase2003class,
  title={Class of algorithms for decoding block codes with channel measurement information},
  author={Chase, David},
  journal={IEEE Trans. Inform. Theory},
  volume={18},
  number={1},
  pages={170--182},
  year={2003},
  publisher={IEEE}
}

@article{wu2006soft,
  title={Soft-decision decoding of linear block codes using preprocessing and diversification},
  author={Wu, Yingquan and Hadjicostis, Christoforos N},
  journal={IEEE Trans. Inform. Theory},
  volume={53},
  number={1},
  pages={378--393},
  year={2006},
  publisher={IEEE}
}

@inproceedings{jin2006probabilistic,
  title={Probabilistic sufficient conditions on optimality for reliability based decoding of linear block codes},
  author={Jin, Wenyi and Fossorier, Marc},
  booktitle={IEEE Int. Symp. Inf. Theor. Proc},
  pages={2235--2239},
  year={2006},
  organization={IEEE}
}

@inproceedings{yue2019segmentation,
  title={Segmentation-discarding ordered-statistic decoding for linear block codes},
  author={Yue, Chentao and Shirvanimoghaddam, Mahyar and Li, Yonghui and Vucetic, Branka},
  booktitle={Proc. - IEEE Glob. Commun. Conf., GLOBECOM},
  pages={1--6},
  year={2019},
  organization={IEEE}
}

@article{liang2022low,
  title={A low-complexity ordered statistic decoding of short block codes},
  author={Liang, Jifan and Wang, Yiwen and Cai, Suihua and Ma, Xiao},
  journal={IEEE Commun. Lett.},
  volume={27},
  number={2},
  pages={400--403},
  year={2022},
  publisher={IEEE}
}

@inproceedings{li2024order,
  title={Order Skipping Ordered Statistics Decoding and its Performance Analysis},
  author={Li, Xihao and Chen, Wenhao and Chen, Li and Li, Yuan and Zhang, Huazi},
  booktitle={IEEE Inf. Theory Workshop, ITW},
  pages={448--453},
  year={2024},
  organization={IEEE}
}

\end{document}